# Effect of ageing on the properties of the W-containing IRIS-TiAl alloy

Alain Couret[1], David Reyes[1], Marc Thomas[2], Nicolas Ratel-Ramond[1], Christophe Deshayes[1], Jean-Philippe Monchoux[1].

1. CEMES, Université de Toulouse, CNRS, 29 rue Jeanne Marvig, BP 94347, 31055 Toulouse, France
2. ONERA/DMAS Université Paris Saclay, 29 Avenue de la Division Leclerc, BP 72, 92322 Châtillon Cedex, France

* Corresponding author: alain.couret@cemes.fr

*Abstract*

The effects of ageing at 800 °C on the properties of the IRIS alloy ($Ti_{49.9}Al_{48}W_2B_{0.1}$) are studied. The initial microstructure of this alloy densified by Spark Plasma Sintering (SPS) is mainly composed of lamellar colonies which are surrounded by γ grains. The evolutions of the alloy strength and creep resistance resulting from this ageing treatment are measured by the related mechanical tests. The microstructural changes are investigated by scanning and transmission electron microscopies and by X-ray diffraction.

The main structural evolutions consist in a shrinkage of the lamellar areas and in a precipitation of $β_0$ phase, which is accompanied by a moderate segregation of tungsten and a decrease of the $α_2$ lamellar width. However, these evolutions are relatively limited and the microstructural stability is found to result mainly from the low diffusivity of tungsten. Conversely, a moderate effect of this ageing treatment on mechanical properties, at room and high temperatures, is measured. Such experimental results are interpreted and discussed in terms of the microstructural evolutions and of the deformation mechanisms which are activated at different temperatures under various solicitations.

## 1. Introduction

With the aim to reach a service temperature possibly above 800°C in aeronautic and automotive engines, an appropriate solution to improve the mechanical properties of TiAl alloys is to add refractory elements, like Nb, Mo, Ta and W, with low diffusivity. Indeed, these elements are known to improve the high temperature strength but also the oxidation resistance of these alloys [1]. A well-known example of this is related to the Nb- and Mo-containing TNM alloys [2]. Moreover, these elements also have the property to stabilize the β/$β_0$ phase, which is soft and ductile at high temperatures, making the forging operations easier to be performed.

The high capability of tungsten to improve the creep resistance of TiAl alloys has firstly been demonstrated through the development of cast ABB alloys by Nazmy and Staubli [3]. Alloys of this family have subsequently been investigated in several studies concerning their microstructure and mechanical properties at high temperatures, with a special attention to their structural stability [4-7]. The positive effect of tungsten addition has been attributed to different processes: reduction of the dislocation mobility, stabilisation of the interfaces, and pinning of dislocations or interfaces by fine $\beta_0$ precipitates. More recently, taking advantage of the powder metallurgy (PM) – Spark Plasma Sintering (SPS) route, the W-containing IRIS alloy ($Ti_{49.9}Al_{48.0}W_{2.0}B_{0.1}$, composition in at.%) has been found to offer a good compromise to achieve a reasonable ductility at room temperature and a high strength at elevated temperature [8-9]. Boron has been incorporated to reduce the grain size, a well-known effect for TiAl alloys which has been recently illustrated in the case of alloys densified by the SPS route [10]. The plastic elongation at room temperature of the IRIS alloy is due to the control of the grain size which results from the presence of both borides and β phase at high temperatures, thus forming obstacles to the α grain growth [11].

Thermal treatments performed in the range 700°C - 1000°C for W-containing TiAl alloys have been found to lead to a number of microstructural evolutions such as the formation of $\beta_0$/B2 phase as, precipitates, coarsened rods or fine needles, the dissolution of $\alpha_2$ lamellae and the growth of γ phase [4;12-16]. Depending on the alloy characteristics and on the applied thermal treatment, these microstructural changes have been associated either to an improvement or to a reduction of the alloy mechanical properties. To gain further understanding of these phenomena, the aim of the present work is to identify whether an ageing treatment such as 500 hours at 800 °C could be deleterious or not on the microstructure and mechanical properties of the IRIS alloy.

## 2. Experimental

The alloy used in this work was densified by Spark Plasma Sintering from an IRIS pre-alloyed powder (Ref A1261) atomised by the ATI company (Pittsburgh, PA, USA). SPS experiments were performed on the Dr Sinter Sumitomo 2080 apparatus (Sumitomo Coal Ming Co. Japan) implemented in the "*Plateforme Nationale de Frittage Flash /CNRS*" in Toulouse, France. The measured dwell temperature of the SPS cycle was 1290°C, which corresponds to a temperature for the material of 1350°C [17]. The chemical composition of the IRIS–SPS billet as measured by Inductively Coupled Plasma Spectroscopy was found to be $Ti_{50.11}Al_{47.92}W_{1.86}B_{0.12}$. The oxygen content as measured by Instrumental Gas Analysis is 1200ppm wt. A single ageing treatment was carried out in this work, which consists of 500 hours at 800°C in air, followed by furnace cooling. This treatment was applied on the as-SPSed billet and then the tensile specimens were machined from these treated materials.

The microstructural investigations and local chemical analyses were performed by scanning electron microscopy (SEM) and transmission electron microscopy (TEM), using four different microscopes. Two SEM were used: a JEOL 6490 equipped with a tungsten emission gun for standard observations and a FEI HELIOS Nanolab 600i field emission gun (FEG) dual focused ion beam device for high resolution investigations. Local chemistries of Ti, Al and W contents were measured by Energy-dispersive X-ray spectroscopy (EDS) in the SEM and TEM and will be given in at. % in the following. Every experimental mean value is the result of at least ten measurements. Boron, the role of which being to limit the grain growth, will not be discussed in this paper. The SEM micrographs were obtained using back-scattered electron mode. A JEOL 2010 TEM equipped with a LaB6 cathode was used for structural investigations and orientation determination whereas the chemical measurements were performed in a Philips CM20 TEM equipped with a field emission gun. The selected dark field technique which was used to determine the orientations of lamellae and precipitates, is described in details elsewhere [18]. Phase identification was also carried out by means of X-ray diffraction using a Bruker D8 Advance diffractometer in Bragg-Brentano geometry (Cu-Kα radiation). Quantitative data were extracted from SEM micrographs by using ImageJ software. The grain size is evaluated by considering the diameter of a circle covering the same area as the grain.

Tensile tests were performed at a constant strain rate of $10^{-4}$ $s^{-1}$ at room temperature (CEMES), 800°C and 900°C (ONERA) and the creep tests were carried out at 800°C under 200 MPa (ONERA).

## 3. Microstructure of the as-SPSed alloy

Fig. 1 shows the microstructure of the as-SPSed IRIS alloy, which is mainly formed of lamellar colonies [11]. At grain boundaries, single-phased γ zones, hereafter called γ borders, consist of several adjacent γ grains. The orientations of these γ grains of the borders have been found identical to those of one of their neighbouring lamellae [11]. Consistently, Fig. 1 illustrates that the two sides of the borders are different. Firstly, from one side where the γ grains and the lamellae exhibit the same orientation, their separation limit is diffuse (yellow arrows) and the γ grains appear to be rather an extended zone of the lamellar colonies. Secondly, from the other side, the boundary forms a clear limit (green arrows) with the other neighbouring lamellar colony. The γ borders are present irrespectively of the lamellar interface orientation with respect to the grain boundary. These borders contain precipitates of $β_0$ phase (white dots), which are generally attached to the well delimited boundary and extend towards the border. It has been found that the orientation relationship of $β_0$ phase with the γ phase does follow the Kurdjumov-Sachs relation $\{110\}β_0//\{111\}γ$ and $<111>β_0//<110>γ$ [17]. A few single-phased γ zones are also observed within lamellar colonies, while also containing $β_0$ precipitates. Similar observations of equiaxed/globular γ grains have been presented in various W-containing TiAl alloys [7,19,20].

One particular quantitative study was conducted on an area containing 100 adjacent grains, covering a total area of 32 495 µm², with a statistical analysis of every grains and borders. The average size of the included grains, lamellar colonies and borders, corresponds to 17.8 µm; that of the lamellar colonies (borders excluded) is 14.1 µm. The single-phased zones (borders and intragranular zones) cover 27% of the surface fraction. The average chemical composition of the γ phase of the borders, measured by SEM-EDX, is 48.7Ti-49.2Al-2.1W in at.%.

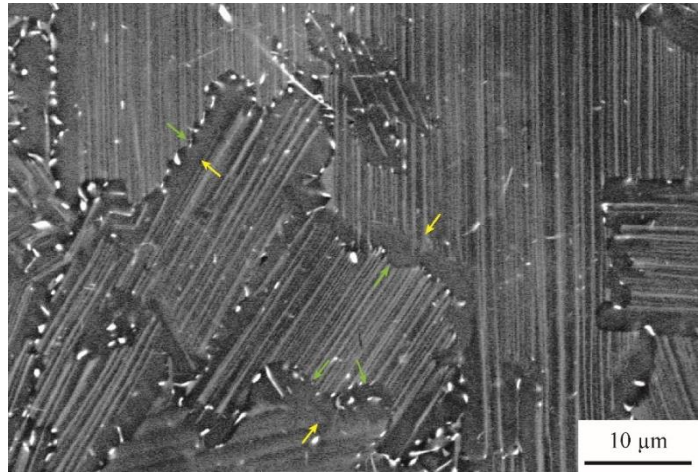

Fig. 1. Microstructure of the as-spsed IRIS alloy.

## 4. Mechanical properties

The mechanical properties of the as-SPSed and aged IRIS alloys are compared at room and high temperatures in Figs 2 and 3. The experimental results are also compiled in Tables 1 and 2.

At room temperature, from the unaged alloy six tensile tests were performed and have led to a yield stress (YS) ranging between 397 MPa and 464 MPa and a plastic elongation at rupture ranging between 0.86 % and 1.89 % [20]. Such a scattering is relatively high for a SPS alloy since the reproducibility in tensile properties is generally very high for SPS-TiAl alloys due to the PM route which enhances the uniformity of microstructures (see for instance Fig. 7b of [22] and Fig. 5a of [23]). The average behaviour of this alloy is represented by the blue curve in Fig. 2 a and by the data given in Table 1. Even if only one tensile experiment was performed on the aged alloy, nevertheless the following differences can be stated between the room temperature behaviour of the as-SPSed and aged IRIS alloy. The ageing has no significant effect on YS but induces a slight drop of the plastic elongation by about 0.2% and of the ultimate tensile strength (UTS) by about 60 MPa. Indeed, one may consider that such a reduction of the plastic strain is significant, as this plastic strain and the maximum strength are concomitantly reduced.

At 800°C and 900°C, a decrease in YS by about 20-25 MPa and in UTS by 55-75 MPa can be emphasized, without change in the general shapes of the curves (Fig. 2b). This behaviour is indicative of a moderate decrease in alloy strength due to this 500-hour treatment.

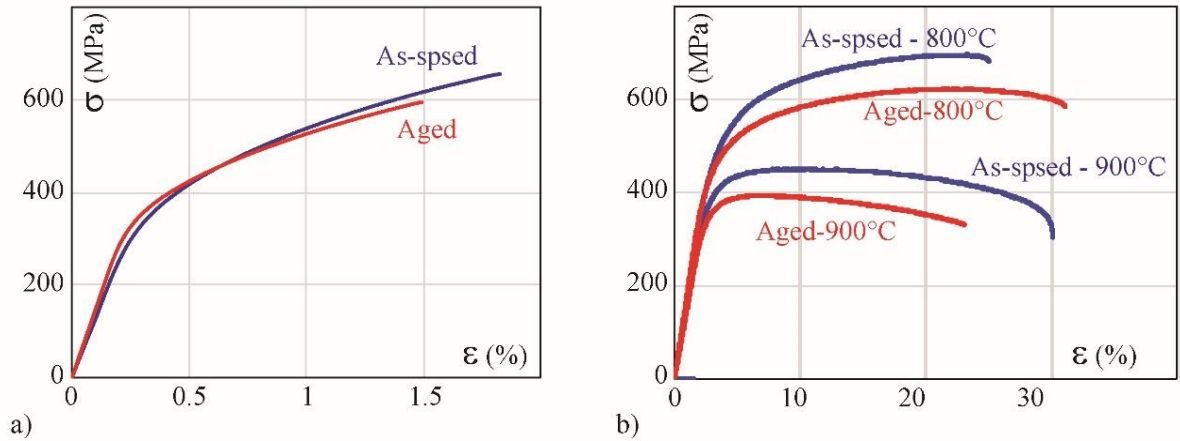

**Fig. 2.** Tensile curves at room temperature (a), 800°C and 900°C (b) of the as-spsed and aged IRIS-TiAl alloys. The temperatures specified on (b) are the deformation temperature.

**Table 1** Results of the deformation experiments at constant strain rate ($10^{-4}$ $s^{-1}$) performed on the as-SPSed and aged alloys at room temperature, 800°C and 900°C.

| Deformation Temperature | Alloy | YS (MPa) | UTS (MPa) | A (%) |
|---|---|---|---|---|
| RT | As-SPSed | 432 | 656 | 1.30 |
|  | Aged | 422 | 594 | 1.08 |
| 800°C | As-SPSed | 420 | 696 | 21.1 |
|  | Aged | 396 | 623 | 27.8 |
| 900°C | As-SPSed | 344 | 450 | 28.2 |
|  | Aged | 322 | 393 | 20.8 |

Fig. 3a shows the creep curves of both unaged and aged IRIS alloys, in comparison with that of a cast GE alloy (GE-C) previously studied at CEMES [24,25]. The latter, which exhibits a microstructure made of a central equiaxed area surrounded by columnar grains, is similar to those currently used in GEnX and LEAP engines by GE and SAFRAN companies [26]. As compared with the GE-C alloy, the lifetime of the as-SPSed IRIS alloy is significantly higher (Tab. 2). Fig. 3a indicate a lower creep resistance of the aged alloy with respect to that of the unaged alloy. This behaviour raises the question of whether the loss of strength is higher at low strain rate (creep) than at high strain rate (tensile tests at constant strain rate). For a better quantification of these differences, the creep properties of these three alloys have been reported on a Larson-Miller diagram using the relationship: $LM : T.(20 \log(t)). 10^{-3}$, where LM is the parameter of Larson-Miller at 1% of strain, T the temperature in kelvin and t the time in hour. From this diagram, the green points and the green curve represent the GE-C alloy for which 27 conditions applied in the range 650°C - 850°C were tested. The points in pink, for the same GE-C alloy, result from 8 tests carried out at the same condition (750°C-150 MPa) that we had studied in detail. This diagram illustrates the creep property scatter for an alloy obtained by the casting route which, however, is certainly expected to be higher than for a PM alloy. The blue marks give the results for four conditions used for the unaged IRIS-SPS alloy (700°C/300MPa, 750°C/200MPa, 800°C/200MPa and 850°C/150MPa), with the blue square corresponding to the 800°C/200MPa condition. The red square corresponds to that of the aged alloy for the same 800°C/200MPa condition. Using this Larson-Miller diagram, it can be emphasized that the drop in creep strength resulting from the exposure of 500 hours at 800°C appears to be relatively moderate, compared to the scatter for the GE-C alloy at 750°C/150 MPa, and keeping in mind the difference in strength between the GE-C and IRIS-SPS alloys. Therefore, this creep property reduction is likely to be related to the loss in YS as observed for high temperature tensile tests at constant strain rate. For both tests, the similarity and the repeatability of the curves in the unaged and aged alloys must be underlined. This is indicative of the fact that, at least for each solicitation, the loss in mechanical strength with ageing is related to a microstructural characteristic that affects the overall mechanical properties.

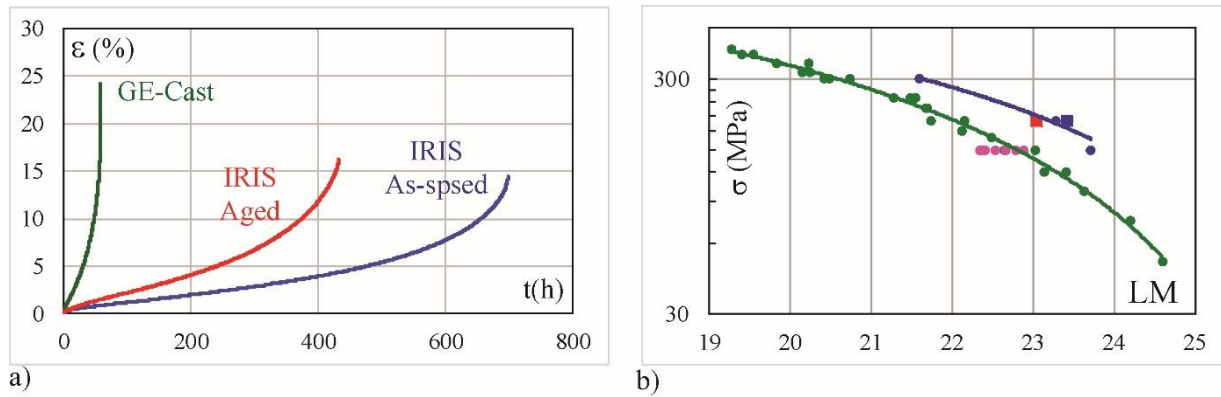

**Fig. 3.** Creep behaviour of the as-spsed and aged IRIS-TiAl alloys in comparison to those of a GE alloy obtained by the cast route. (a) creep curves at 800°C-200MPa. (b) Larson-Miller plot at 1% of strain (LM is the Larson-Miller parameter - see text for details).

**Table 2.** Results of the creep experiments performed on different alloys (see the text for details).

|  | Solicitations | $V_{min}$ (s$^{-1}$) | $t_{rup}$ (h) | $t_{1\%}$(h) | $t_{5\%}$ (h) |
|---|---|---|---|---|---|
| IRIS as-SPSed | 800°C/200MPa | 2.8 10$^{-8}$ | 669 | 67.5 | 473 |
| IRIS aged | 800°C/200MPa | 4.5 10$^{-8}$ | 431 | 29.5 | 238 |
| GE-C | 800°C/200MPa | 3.9 10$^{-7}$ | 58 | 4.2 | 29.7 |
|  |  |  |  |  |  |
| IRIS as-SPSed | 700°C/300MPa | 3 10$^{-9}$ | 4052 | 154 | 2838 |
| IRIS as-SPSed | 750°C/200MPa | 5 10$^{-9}$ | 3138 | 569 | 2641 |
| IRIS as-SPSed | 850°C/100MPa | 6 10$^{-8}$ | 260 | 13 | 146 |

## 5. Microstructure of the aged alloy

As shown in Fig. 4, the microstructure of the aged alloy looks very similar to that of the non-aged one. Quantitative measurements of the surface fractions of the different constituents of the microstructure are given in Table 3. The average size of the grains, lamellar colonies and borders included, are unchanged during the thermal exposure. In contrast, these measurements reveal an increase of the surface fraction of γ zones at the expense of the lamellar colonies. In the author opinion, the measured variation of 10% is over the uncertainty of this kind of measurements. Accordingly, the measured variation is thus significant for us, which should be related to a shrinkage of the lamellar colonies, as illustrated by the areas encircled in yellow dashed lines in Fig. 4. This phenomenon seems to occur at the extremities of the lamellar colonies, close to the grain boundary, thus leading to an increase of the borders. The $\beta_0$ precipitation is also observed to be increased in the aged alloy, as previously mentioned in [16,209]. The precipitates located in the γ borders are mainly linked to the clearly formed boundary separating the border from one neighbouring lamellar grain. Another microstructural change corresponds to the occurrence of black strips at grain boundaries (some examples are marked by yellow arrows in Fig. 4), which are mostly located at the side of the γ border. Similar black strips were previously evidenced in the same alloy aged during 8200 hours at 750°C [27]. In the following, quantitative measurements of the physical and chemical characteristics of the various zones of the microstructure will be presented.

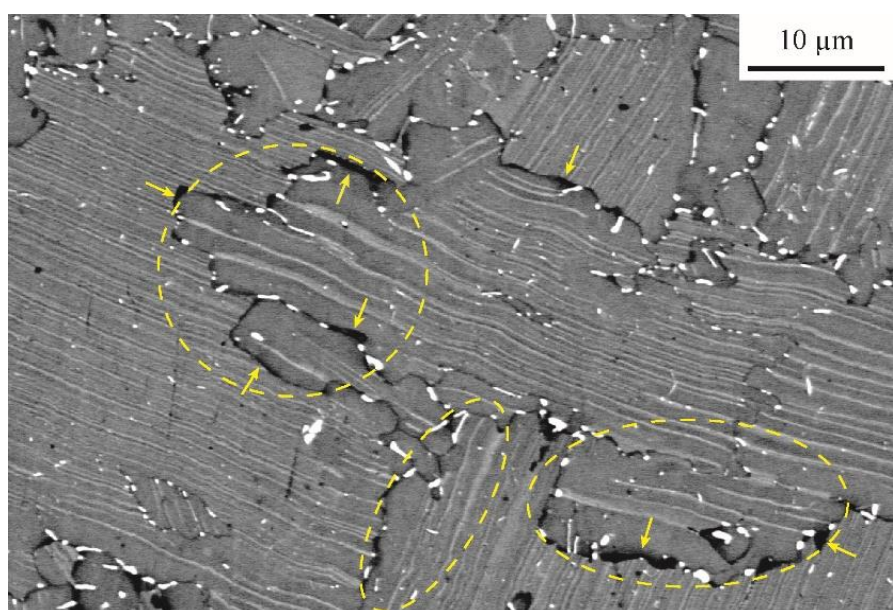

**Fig. 4.** Microstructure of the aged alloy. Yellow arrows indicate black strips at the grain boundaries and yellow encerclements mark zones where the lamellar microstructure is partly vanished.

**Table 3.** Characteristics of the microstructures of the as-SPSed and aged alloys. For the average grain size, the standard deviation are given in brackets.

|  | Total areas (µm²) | Average grain sizes (µm) | Lamellar colonies | | Surface fraction of γ zones (%) |
|---|---|---|---|---|---|
|  |  |  | Average size (µm²) | Surface fraction (%) |  |
| As-SPSed | 32 495 | 17.8 (9.8) | 14.1 (10,1) | 73 | 27 |
| Aged | 33 302 | 17.8 (10,3) | 11.2 (10,2) | 63 | 37 |

*5.1. $\beta_0$ precipitation in the borders*

The chemical compositions of precipitates at the borders were measured by both EDS-SEM and EDS-TEM, which have led to similar results. Average values are given in Table 4. In the as-SPSed and aged alloys, the W content is of the same order of magnitude, about 15 at. %. A slight enrichment in W is detected after the thermal exposure. A reduction of the standard deviation is also measured, which may indicate that the precipitates tend to approach their equilibrium composition. However, due to the small precipitate size with respect to that of the interaction volume between the electron beam and the material from which X-rays are generated, these evolutions will not be further discussed. A W content of about 15% in the $\beta_0$ phase is consistent with that performed previously by Yu et al. [28], which was supported by electronic and energy calculations based on first-principles [29]. These results also agree with the partition coefficients given in [6,30].

**Table 4.** Chemical compositions of the $\beta_0$ precipitates located at the $\gamma$ borders of the as-SPSed and aged alloys. For the chemical contents, the standard deviations are given in brackets.

|  | Al | Ti | W |
|---|---|---|---|
| As-SPSed | 37.6 (4.5) | 49.6 (3.9) | 12.5 (3.1) |
| Aged | 34.8 (3.2) | 48.5 (2.8) | 16.5 (1.3) |

A quantitative study of the number of precipitates located at the borders has been performed. The measurements were repeated in three zones of both unaged and aged conditions (Tab. 5). The images (Fig. 5a) were binarized to isolate the precipitates (Fig. 5b). Note that Fig. 5a only shows a quarter of the area which was investigated, the size of which was 128 µm x 87.6 µm = 11 213 µm² for each measurement. The black areas larger than 1 pixel (0.1x0.1 µm²) were then automatically counted and their surfaces were calculated using ImageJ software. Results appear to be very reproducible from one zone to the others. During ageing, the precipitate number is approximately multiplied by 3.5, and their average surface by 2.5, thus leading to a multiplication of the surface fraction by a factor of 9.

For the aged sample 2 where 1719 precipitates with an average size of 0.12 µm² were measured, those contained in the borders were separated from those located at lamellar colonies. For this purpose, the lamellar colonies were delimited to extract the borders (Fig. 5c) and the image was binarized (Fig. 5d). A number of 956 precipitates, nearly half of the total number, were found in the borders; their average size being 0.18 µm². Consequently, the precipitates of the lamellar colonies should have an average size of approximatively 0.06 µm². This result confirms that the vast majority of the largest precipitates are located at the borders.

**Table 5.** Counting of the number and size of precipitates located at the $\gamma$ borders of the as-SPSed and aged alloys.

|  | Specimens | Numbers | Average Surface (µm²) |
|---|---|---|---|
| **As-SPSed** | 1 | 442 | 0.04 |
|  | 2 | 526 | 0.05 |
|  | 3 | 488 | 0.05 |
|  | **Average** | **485** | **0.05** |
| **Aged** | 1 | 1677 | 0.14 |
|  | 2 | 1719 | 0.12 |
|  | 3 | 1665 | 0.14 |
|  | **Average** | **1687** | **0.13** |

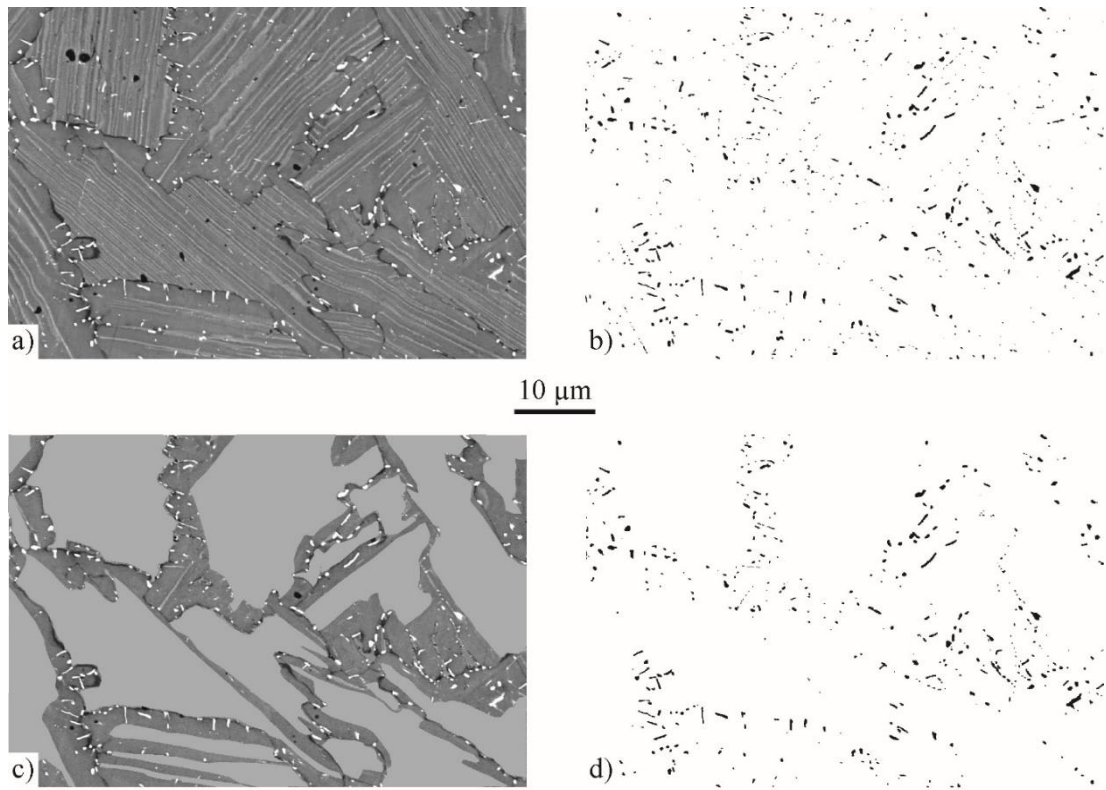

**Fig. 5 .** Example of a quantitative study of precipitates situated in the borders (specimen 2 of the aged alloy – Tab. 5). (a) SEM micrograph; (b) Binarized image of (a) with the precipitates in black; (c) extraction of the borders; (d) Binarized image of (c).

*5.2. Precipitation in the lamellar colonies*

In lamellar colonies of the aged material, TEM allows to show some precipitates which cannot be seen with the JEOL-6490-SEM (Fig. 4). The technique of selective dark field in the TEM has been used to study these very small precipitates, in particular to determine their structure, characteristics and orientations. Fig. 6a, taken under an inclination of the interface plane at 60°, shows a general view of several families of precipitates. Fig. 6b, in bright field mode, with the interface plane normal to the observation plane, displays the precipitate arrangement in a series of thin lamellae (S1), which are alternatively of $\gamma$ and $\alpha_2$ phases, this point being described in detail below (section 5.4). Fig. 6 b shows that the precipitates are spread in the $\alpha_2$ lamellae and that a higher number of precipitates are located in the two lamellae bordering this series than in those located in the middle. For the area surrounded in dashed blue line in Fig. 6a, the series of dark field micrographs taken with a diffracting vector illuminating the precipitates (Figs. 6 c to f) enables us to distinguish these various families, which are elongated along different directions. As particularly evidenced in Figs. 6 c and d, the precipitates often have finite and straight limits, which should be located in a plane parallel to the interface plane since their two extremities are separated by the apparent width of the interface plane. From our observations of many precipitates under different inclinations, we were not able to situate the directions of theses limits along specific crystallographic directions of the $\alpha_2$ lamellae without ambiguity. This might be explained by the effect detected by Zhang *et al.* [31] who have observed that $\beta_0$ precipitates grow in $\alpha_2$ grains along invariant lines resulting from a strain homogenization process. Fig.6 e was taken with the interface plane almost upright (inclined at 7 °). This figure shows that these precipitates are not planar and located in the interface plane but, spread out of this plane.

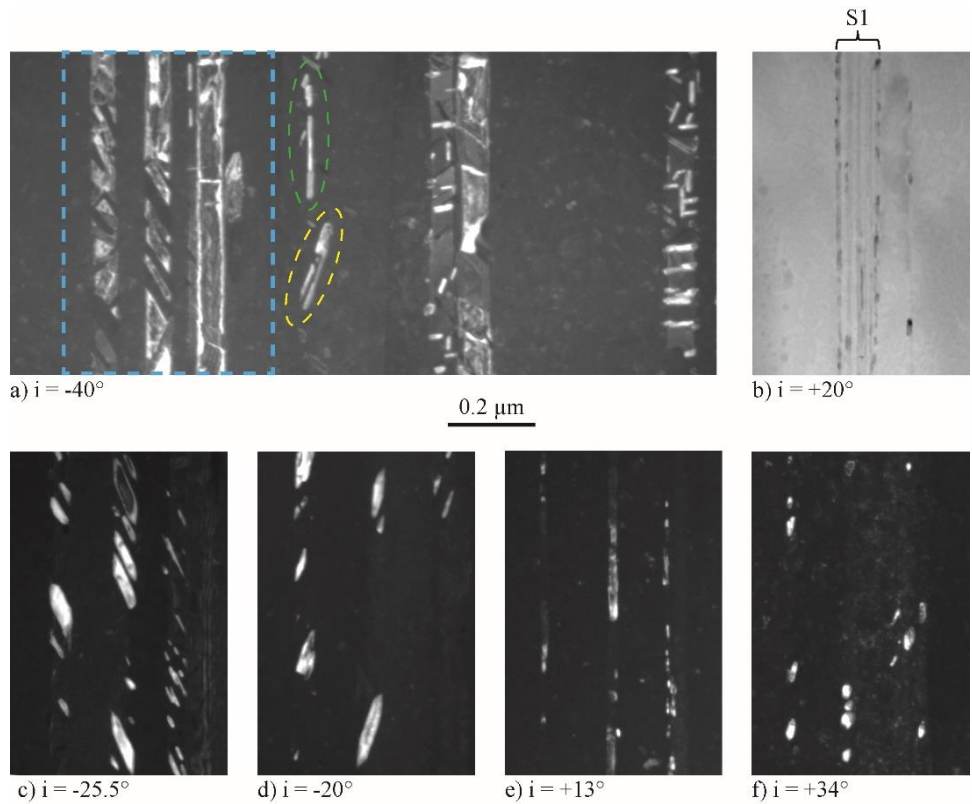

**Fig. 6.** Micrographs of the precipitates in lamellar zones. i gives the inclination angle of the sample when the picture was taken, whereas for this grain the interface plane was inclined at + 20° under zero tilt. (a) general view in dark field mode under an inclination of +40 °. (b) micrograph in bright field mode with the interface plane edge on of the series S1 of thin lamellae characterised in section 5.4. (c) to (f) dark field micrographs of the area surrounded in blue on (a).

Because of their very small size, it was not possible to determine the precipitate structure and orientation by analyses in electronic precession diffraction or convergent beam electron diffraction modes. Fig. 7 presents the study of one family of precipitates illustrated in Fig. 6a (surrounded in green), using the selective dark field technique. The precipitates are in contrast with five diffracting vectors which are all compatible with the B2 structure. Using classical stereographic methods, their orientations were determined from the tilt angles and from the diffracting vector directions given by the diffraction patterns. It is represented by the stereogram (Fig. 7 f), for which the projection plane is the thin foil plane under zero tilt. Comparing this orientation to those of the γ and $\alpha_2$ lamellae of the neighbouring lamellae, a coincidence relation can be put forward, for which the plane of interface of the lamellar structure is parallel to the plane (110) of the $\beta_0$ phase. Moreover, a dense direction of $\beta_0$, [111], is found to be parallel with a direction of type $<11\bar{2}0>$ of $\alpha_2$ phase. These results concerning the crystallographic structure and orientation of these $\beta_0$ precipitates are fully consistent with those previously obtained in a Ti-47Al-2Nb-1Mn-0.5W-0.5Mo-0.2Si cast alloy aged in the range 760 C - 1050°C [14]. Clearly, the studied precipitates keep a high aspect ratio under different inclinations, which is consistent with a needle shape. Many precipitates which are shown in Fig. 6a do not have this needle shape but seem to be extended in a plane parallel to the interface plane of the area. The precipitates are likely to grow firstly with this needle shape and subsequently spread in other directions. Accordingly, precipitates with needle shape and planar spreading are observed for the same family (Figs. 6 c and d). The same kind of structure and orientation determination of precipitates was successfully repeated for that surrounded in yellow in Fig. 6 a.

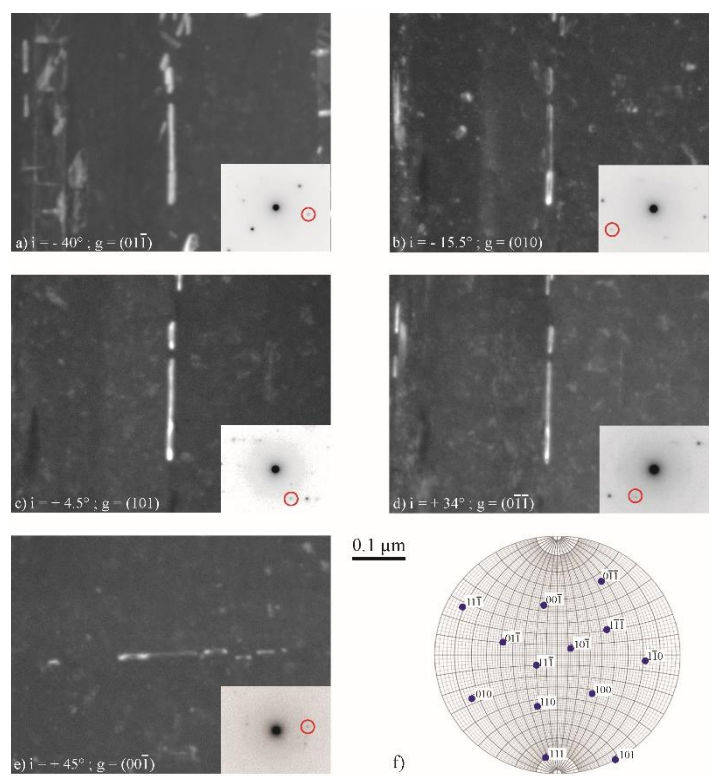

**Fig. 7.** Micrographs in dark field mode of one family of $\beta_0$ precipitates. For each micrograph the diffraction pattern in bright mode is reported, with the red circle indicating the used diffracting vector for the dark field. For recording the micrograph (e), the specimen was rotated of 90° around the direction of the electron beam. (f) stereogram under zero tilt representing the orientation of this family of precipitates.

### 5.3. Evolution of the chemical composition in the borders

From the SEM micrographs (Fig. 8a), thin black bands are connected to a grain boundary. They are rather located at the side of the γ phase than on that of the lamellar colonies. The EDS W mapping shows in Fig. 8b that these zones are depleted in W. Table 6 gives the chemical composition of the γ borders in the unaged and aged alloys, with a distinction for the latter between the normal and W depleted zones. The γ phase is not affected by the exposure of 500 hours at 800°C except in such W depleted zones. Figs. 4 & 8 apparently show that this W depletion is accompanied by an elongation of $\beta_0$ precipitates beyond the limit of this zone, and this mainly in a direction perpendicular to the grain boundary.

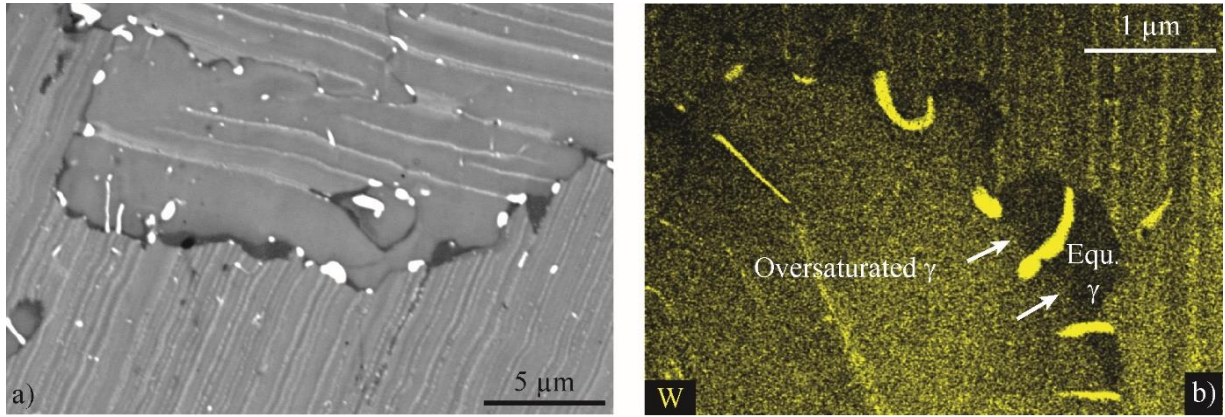

**Fig. 8.** Thin black bands in a γ border. (a) SEM micrograph; (b) STEM-EDS map of W. Arrows indicate the sharp transition between the oversaturated and equilibrated zones of the γ phase.

**Table 6.** Chemical composition of the borders in the as-SPSed and aged alloys. For the chemical contents, the standard deviations are given in brackets.

|  |  | Al | Ti | W |
|---|---|---|---|---|
| As-SPSed | γ zones | 49.2 (6.6) | 48.7 (5.6) | 2.1 (0.6) |
| Aged | γ zones | 49.2 (5.5) | 48.7 (5.5) | 2.05 (0.4) |
|  | W depleted zones | 49.3 (1.6) | 50.0 (1.7) | 0.7 (0.2) |

### 5.4. Transformations in the lamellar colonies

Fig. 9a displays a TEM micrograph of a lamellar colony for an aged sample, with the interfaces upright and parallel to the tilt axis of the TEM specimen holder. The phases and orientations of every lamella which are specified in Fig. 9b were determined using the selective dark field technique. It is recalled that in lamellar TiAl alloys, γ lamellae precipitate in α grains along the (0001) basal plane and exhibits six different variants differing by the stacking sequence along the normal of the {111} interface (as …ABCABC… or …ACBACB…) and/or by the relative orientations of the <110] monoatomic row in the planes parallel to the interface [18]. This results in the formation of three different types of γ/γ interfaces in addition to the γ/$α_2$ interface in the lamellar structure: ordered domain, true-twin and pseudo-twin interfaces. Only two dark field micrographs among the seven which were necessary for a full characterization of this area are reported here (Figs. 9c and d). Fig. 9c points out the $α_2$ lamellae and Fig. 9d those of one γ orientation (labelled O1). Fig. 9e shows the chemical composition profiles of the three major elements, integrating over the whole area of Fig. 9a. For comparison, Fig. 9f displays the composition profiles in an unaged sample recorded in the same experimental conditions, with the insert giving a view of the microstructure. For the area in Fig. 9a, Fig. 10 shows the composition profiles at three locations separated by 700 nm along the interface plane, with an integration width of 100 nm.

The following experimental results can be highlighted from this TEM study:

- This lamellar zone contains 15 lamellae (marked L1 à L15 on Fig. 9b) and two series of thin lamellae (marked S1 and S2) which are formed by a perfect alternation of γ and $α_2$ lamellae. As often observed in lamellar TiAl alloys [32], one γ orientation is predominant (O1 in this example) and thin $α_2$ lamellae are often present between two γ lamellae which are pseudo-twin related (L6 between L5 and L7, L10 between L9 and L11, and L12 between L11 and L13), which means that their orientation relationship involves both a fault of order and a change of the stacking sequence of interface planes [18]. The formation of wide γ lamellae predominantly oriented with the same orientation and that of series of fine lamellae result from the activation of two different lamellar transformations activated at high and lower temperatures, respectively [32].

- In the aged alloy (Fig. 9e), an enrichment in W is measured at the $α_2$ lamellae (see for example L4 and L6, the positions of which are marked in figures) which is more irregular and more pronounced than in the as-SPSed alloy in which these W peaks at $α_2$ lamellae are regular and approximately equal to 3 at. % (Fig. 9f). This amount of 3 W at. % in the $α_2$ phase is consistent with the measurement of Yu et al. [28]. The local profiles displayed in Fig. 10 show that high W peaks (encirclements in blue) or weak peaks (zones encircled in red) could be indifferently measured at the positions of $α_2$ lamellae, within a small area analysis. The highest peaks of W reach 10 at. %. Let us assume that the W enrichment which occurs during the ageing treatment is concentrated in the $β_0$ precipitates, as described in section 5.2 in a similar manner as the precipitates in the borders. Hence, the abrupt variations of the W content along $α_2$ lamellae could be associated to the irregular presence of $β_0$ precipitates.

- Looking at these fine and isolated $α_2$ lamellae, irregular depletions in aluminium and enrichments in titanium and tungsten are measured. The higher W enrichments correspond to the larger Al depletion (L4 and L6 compared to L10 and L12 on Fig. 9e). A Ti enrichment is measured for L6 but is null (L10) or quasi-null (L4 and L12) at some other $α_2$ lamellae. In comparison, in the non-aged alloy (Fig. 9f), the variations of the Al, Ti and W contents are more regular and enrichment in Ti is always present at $α_2$ lamellae, thus clearly highlighting the presence of $α_2$ lamellae (in dark in the insert of Fig. 9f). It thus appears that, simultaneously to the $β_0$ precipitation, irregular decreases in the width of $α_2$ lamellae occur during the thermal treatment. However, very thin $α_2$ lamellae and interfaces are still present at these areas (Fig. 9c), thus indicating that $α_2$ lamellae were not fully dissolved in these typical lamellar areas.

- At the series of fine lamellae (encirclements in green in Fig. 10), the W enrichment is less pronounced and mainly located at the external lamellae of the series (for the clearest example, see series 1 in profiles 1 and 3 of Fig.10 b and d).

- Even if the resolution on the chemical composition is relatively limited, these profiles indicate that, despite the segregation in $β_0$ precipitates, a dramatic reduction in the W content for the γ lamellae is not observed. EDS-TEM measurements in the γ lamellae have led to nearly the same values: 1.89 in the as-SPSed alloy and 1.85 in the aged one.

As a summary, we have observed in the lamellar zone the formation of fine $β_0$ precipitates in $α_2$ lamellae which is accompanied by a partial dissolution of these $α_2$ lamellae and by the segregation of tungsten atoms coming from the γ lamellae. The W content in γ lamellae is nearly unchanged.

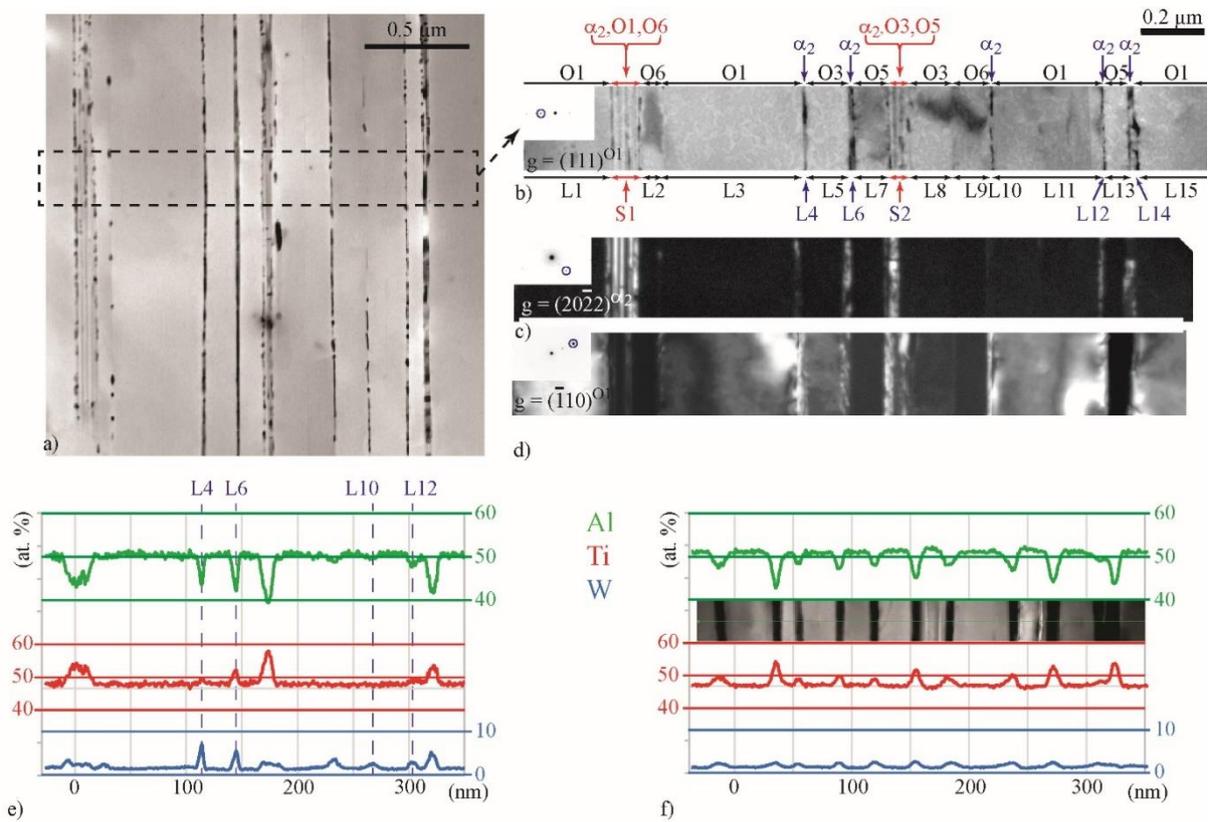

Fig. 9. Study of the lamellar microstructure in the aged alloy. (a) view with the interfaces upright and parallel to the tilt axis of the TEM specimen holder, (b) phases and orientations of the lamellae, (c) – (d) dark field images, (e) chemical composition profile of the three major elements integrated over the whole area; (f) for comparison, chemical composition profiles in a non-aged sample. Green, red and blue profiles correspond to aluminium, titanium and tungsten elements, respectively.

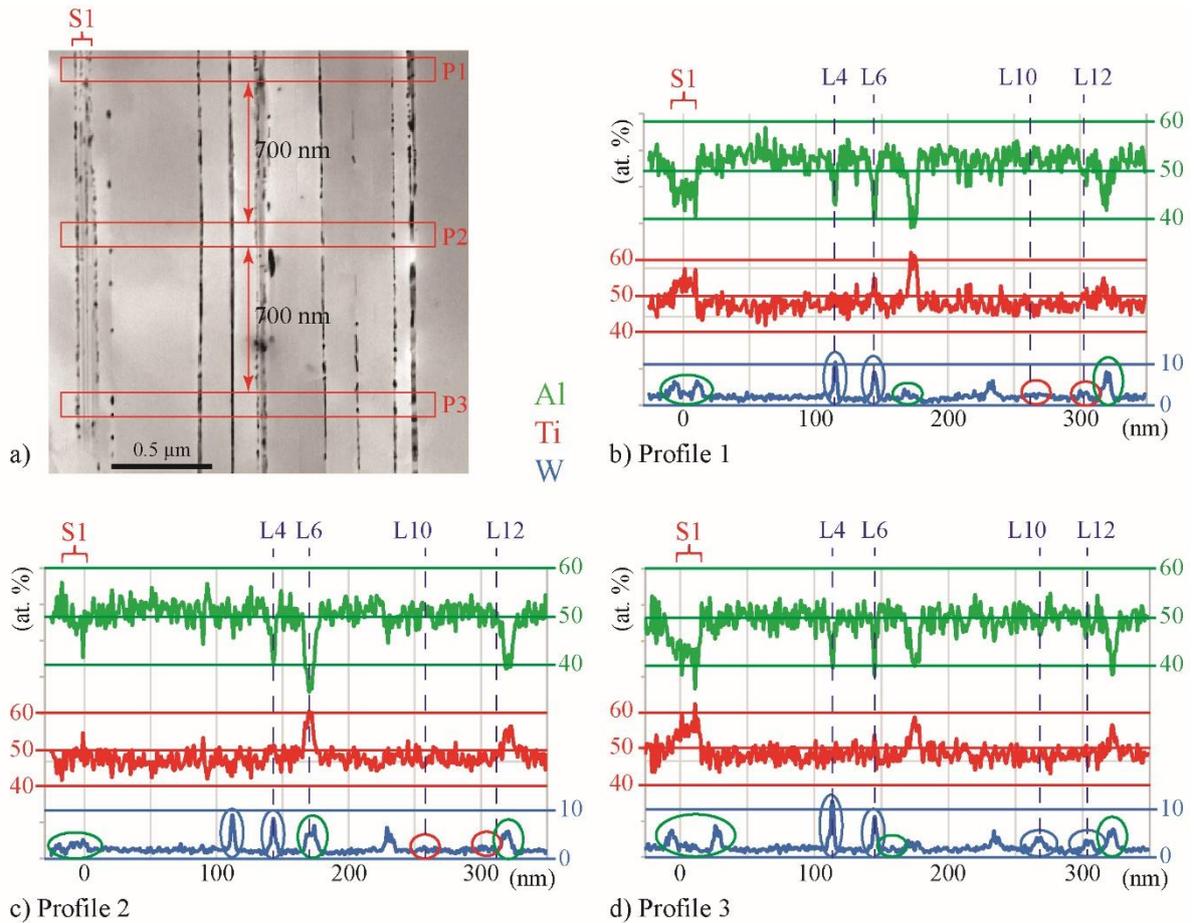

Fig. 10. Study of the local chemical composition of the lamellar microstructure in the aged alloy. (a) view of the zone, (b) to (d) local chemical composition profiles of the three major elements for the three areas P1, P2 and P3 marked in red on (a). Green, red and blue profiles correspond to aluminium, titanium and tungsten elements, respectively.

*5.5. Evolution of the volume fraction of phases*

Quantitative phase analysis was performed using Rietveld refinement on X-ray diagrams obtained on as-SPSed and aged specimens. On both specimens, X-ray diagrams show the presence of the three model phases: $\gamma$ – TiAl type tetragonal structure, $\alpha_2$- Ti$_3$Al type hexagonal structure, and $\beta_0$ type cubic structure. The chemical compositions of these model structures are different from those determined from EDX measurements. For this reason, the chemical compositions of the three model phases used in the Rietveld refinement model were adjusted to fit the chemical compositions measured by EDX in this study and reported in Table 4. The occupancy factors of atoms were adjusted and Ti/Al substitution on the Ti sites have been considered. The results of the quantitative phase analysis are given in Table 7.

Upon annealing, XRD results show that the fractions of $\gamma$ and $\alpha_2$ phases slightly decrease while the fraction of $\beta_0$ increases. This result is in full agreement with the microstructure observations, which have shown a larger content of precipitates in the annealed specimen, accompanied with a decrease in size of the lamellar grains and of the $\alpha_2$ lamellae. The decrease in the $\gamma$ phase fraction results from the fact that the $\beta_0$ precipitation is more pronounced than the $\alpha_2$ dissolution in lamellar zones, due to the low initial fraction of the latter phase. The thermally activated migration of W atoms from γ and $α_2$ phases seems to favour the formation of the precipitate phase.

Table 7. Fraction of phases (at%) determined using Rietveld refinement.

| Phase | As-SPSed | Aged |
|---|---|---|
| γ | 89.6 | 88.2 |
| $α_2$ | 9.2 | 8.3 |
| $β_0$ | 1.2 | 3.5 |

## 6. Discussion

During the ageing treatment of 500 hours at 800 °C for the IRIS alloy, the first change in the microstructure is the precipitation of $β_0$ phase in both single phased γ zones and lamellar colonies. In γ borders, growth of existing precipitates and nucleation of new ones have been evidenced, thus leading to the multiplication by a factor of nine of the volume fraction of $β_0$ phase. The main consequence of this precipitation is the formation of W-depleted thin bands situated at grain boundaries. In the lamellar colonies, needle shaped and planar $β_0$ precipitates are formed in the $α_2$ lamellae, leading here again to W segregation. However, the full dissolution of $α_2$ lamellae is not observed consistently with the limited size of these precipitates. Segregation of W atoms in precipitates comes from the partially dissolved $α_2$ phase and from the neighbouring γ lamellae.

On the one hand in the aged alloy, counting the precipitates and measuring their size (section 5.1) from SEM micrographs, have resulted to a mean number and size of precipitates of 1687 and 0.13 μm² in an area of 2 480 μm². This corresponds to a surface fraction of $β_0$ phase of 1.95% in the aged material, in comparison to 0.21% obtained by the same method in the as-SPSed alloy. On the other hand, the volume fraction of $β_0$ phase has been measured to be 3.5 and 1.2 in the same alloys by XRD (section 5.5-Tab. 7). In the case of SEM investigation with the JEOL-6490-SEM, the smallest precipitates, particularly those in the lamellar areas, could not be revealed, which can partially explain the discrepancy between the measurements obtained with these two experimental methods. In order to obtain a rough estimation, we will take as approximate values, 1% and 3% as $β_0$ phase contents in the unaged and aged alloys respectively. Taking a W content of 15 at. % in this $β_0$ phase, it is found that only 7.5 % and 22.5 % of the W atoms which were initially incorporated in the IRIS alloy, were found to segregate in the $β_0$ precipitates of the unaged and aged alloys. Therefore, whatever the alloy condition, aged or non-aged, a large fraction (more than 75%) of the incorporate tungsten atoms is still present in the γ phase for a hardening purpose. These evaluations are consistent with the data obtained in the present work by XRD (Tab. 7), i.e. a γ phase presents in nearly 89.6 % and 88.2 % with a W content of 2 at. % and an $α_2$ phase present in nearly 8.2 % and 8.3% with a W content of 3 at. %, respectively in the aged and non-aged alloys.

The thin bands depleted in W, which form at the lamellar grain boundaries after ageing, are indicative of diffusion and segregation phenomena (Fig. 8 and Table 6). The driving force for these mechanisms is probably the supersaturation of W in the γ phase. During the SPS cycle, the material is brought to the single α phase field, in which the W solubility is quite high, ranging from 1.06 at. % [33] to 2.4 at. % [30] depending on the authors. Then, during subsequent cooling, the α phase transforms into lamellar colonies, which is mainly composed of γ phase (> 90 %), in which the W solubility is significantly lower than in the α phase: values of 0.62 at. % [33] and 1.06 at. % [30] are reported by the same authors. This results in supersaturation of W in the γ phase. Indeed, the onset of W precipitation has actually been observed during SPS processing [9,23]. However, due to the rapid of SPS processing, the desaturation process of the γ phase is only partial and there should be a tendency for W to segregate into precipitates during ageing subsequently to SPS processing. The observation of an increase in the number and size of β precipitates at boundaries between lamellar grains after ageing (Table 5), which is accompanied by a localized W depletion at these locations (Fig. 8), support this assumption. More precisely, the desaturation of the γ phase occurs by nucleation and growth of β precipitates at the boundaries between lamellar colonies, which are the most favorable sites in terms of interfacial energy, by a classical heterogeneous nucleation mechanism. The mass transport of W involved in this mechanism occurs by diffusion, firstly in the γ phase, then in the grain boundaries. However, the lateral extension of the W-depleted zone in is rather limited (≈ 1 μm, Fig. 8), which suggests that the desaturation in W of the γ phase is difficult at 800°C. The dissolution of W in the γ phase is therefore considered to be minor. In the lamellar areas, the desaturation driving force for the precipitation is accompanied by the one coming from the $\alpha_2$ dissolution. An intriguing observation can however be made in the W-depleted zones (e.g. Fig. 8b): the transition between the supersaturated γ (in bright) and the equilibrated γ (in dark) is rather abrupt (see arrows in Fig. 8b). This is unexpected, because volume diffusion phenomena generally induce smoother compositional variations. This point is not fully elucidated yet, but a potential interpretation could be a strong concentration dependence of the solute diffusion coefficient of the W atoms. Indeed, mathematical analyses show that, in the case of strong diffusion coefficient variations with concentration, the concentration profiles exhibit abrupt transitions [34].

In the lamellar colonies, the W enrichment at former positions of $\alpha_2$ lamellae is firstly originated from W atoms within the $\alpha_2$ lamellae, as a consequence of the partial dissolution of the fine $\alpha_2$ lamellae. In this case, the fine lamellar zones would enhance the diffusion of W atoms compared to the large γ zones. However, at the series of fine lamellae, the W enrichment is less pronounced and mainly located at the external lamellae of the series (section 5.4 – Fig. 10 and 11). This observation is consistent with the observation of a high number of precipitates in the lamellae bordering this specific series (section 5.2 – Fig. 6b). This suggests that, in addition to the segregation coming from the $\alpha_2$ lamellae dissolution, the W atoms segregating in the precipitates are at least partly coming from γ lamellae, over a distance larger than the width of the fine γ lamellae of the series. Indeed, for the $\alpha_2$ lamellae inside the series, their bordering γ lamellae are too fine to provide enough tungsten atoms.

It has been measured that the ageing treatment results in a reduction of the surface fraction of the lamellar microstructure from 73% to 67% through the expansion of the γ borders at the periphery of the lamellar colonies. The associated reduction of the volume fraction of $\alpha_2$ phase is only of 1% due to the low volume fraction of this phase in these lamellar colonies. Such a diminution of the fraction of lamellar areas might however be assumed to have dramatic consequences on the mechanical properties since the number of interfaces is then reduced. The fraction of γ phase globally decreases of 1% as a consequence of the balance between different mechanisms: the precipitation of $\beta_0$ phase, the decrease of the $\alpha_2$ lamella width and the diminution of the lamellar areas. Except in the zones directly affected by some microstructural changes, as the W-depleted bands in the borders or the lamellae in which lamellar $\beta_0$ precipitates have nucleated, the chemical composition of the phases do not vary so significantly.

For all types of solicitations, 20°C, 800 °C and 900 °C in tension at constant strain rate of $10^{-4}$ s$^{-1}$ or during creep at 800 °C under 200 MPa (section 4), the IRIS alloy strength is moderately affected by this ageing treatment of 500 hours at 800 °C and by the microstructural changes. Previous works on W-containing TiAl alloys have resulted in opposite effects of ageing treatments on mechanical properties: Beddoes *et al.* [13] and Zhu *et al.* [16] noted a reduction of the instantaneous primary creep strain and of the creep strain rate interpreted as an effect of the $\beta_0$ precipitation at interfaces whereas Munoz-Morris *et al.* [6] measured a reduction in YS at 20 °C and 700 °C, which was attributed to the dissolution of the $\alpha_2$ phase. In the author's opinion, the interpretation of these ageing effects requires deep understanding of the microstructure evolutions versus ageing time and of the deformation mechanisms activated under the various applied solicitations. The deformation mechanisms activated at room and high temperatures in the SPS-IRIS alloy have been investigated elsewhere by the present authors [9,35]. Indeed, it has been shown that, at high temperatures, the interaction between dislocations moving by glide or climb processes and atoms of tungsten is the controlling phenomenon for the alloy strength. At room temperature, the IRIS alloy is not among the stronger TiAl alloys [9] due to its relatively high proportion of single-phased areas (borders) in which ordinary dislocations glide and twins propagate. YS is comparable to that of a GE alloy densified by SPS [36] which confirms that the role of tungsten is not prevalent at room temperature. A significant part of the ductility loss measured in the aged alloy (0.22 %) could be attributed to the increase of the mean free path between obstacles for twins in the $\gamma$ phase with the reduced size of lamellar areas: the longer this distance, the larger the number of piled-up twin dislocations at given stress, the higher the associated internal stress leading to the sample failure. At 800°C under high strain rate, the deformation results from the glide of ordinary dislocations controlled by a frictional force enhanced by the interaction between dislocations and tungsten atoms and from the crossing of $\gamma/\gamma$ and $\gamma/\alpha_2$ interfaces by the mobile dislocations [35]. Here again, the diminution of the interface number through that of the lamellar zones should slightly reduce the alloy strength. However, the effect of W atoms on the intrinsic mobility of gliding dislocations [37], is still present and preserve the IRIS alloy of a drastic loss of strength. During creep at 800°C under low stress, the dislocation propagation occurs by climb of [001] and ordinary 1/2<110> dislocations in the $\gamma$ phase which is controlled by the low diffusivity of tungsten atoms [35,38]. The creep properties of the aged alloy are therefore moderately affected because a suitable amount of tungsten is maintained in this $\gamma$ phase. The measured reduction of the alloy strength is postulated to be resulting from the increase in the volume fraction of $\gamma$ phase up to 37% (Tab. 3) due to the dissolution of the lamellar structure. Indeed, the lamellar structures are well known to be more creep resistant than single-phased or even duplex ones [1].

As a summary, for the three situations investigated in the present work (high strain rate at RT and HT, and creep), the strength of the IRIS alloy is mainly preserved after the ageing treatment of 500 hours at 800°C because it is controlled by the tungsten atoms distributed in the $\gamma$ phase, which have weakly segregated. Conversely, the moderate drop of strength results from the expansion of the single phased $\gamma$ areas, arising from the dissolution of the lamellar counterparts.

Consequently, one way to reduce the loss of creep resistance due to ageing at 800°C, should be towards stabilising the lamellar structure in the as-SPSed IRIS alloy. Incorporation of a small amount of silicon or carbon is a possible solution and an objective for further investigations. Other alternatives under discussion concern the decrease in aluminium content to 47 at. % or in tungsten content to 1.5 at. %, thereby reducing the driving force for the $\alpha_2$ lamellae dissolution, directly for the first case and indirectly for the second one, through the limitation of $\beta_0$ segregation. Now, the question of the stability of the lamellar structure of the as-SPSed IRIS alloy during long - term ageing of about 10 000 hours which correspond to the operation time of turbine blades in aero-engines is still open and will be further investigated, in a context where ABB23 alloys have been found to be unstable during such long processing times [7].

## 7. Conclusion

In the present work, the evolution of the microstructure and of the mechanical properties of the SPS - IRIS alloy during an ageing treatment of 500 hours at 800 °C has been studied. This ageing treatment was found not to reduce drastically the mechanical properties of the alloy. At room temperature, YS is not affected, but the plastic elongation at rupture is decreased by 0.22% in absolute values. At 800 °C and 900 °C, during deformation at constant strain rate, a slight decrease in YS of less than 25 MPa can be emphasized. The creep properties under 800 °C and 200 MPa are moderately affected. This exposure leads to a precipitation of $\beta_0$ phase in the single-phased $\gamma$ areas and in the lamellar colonies, which in turn induces the formation of W-depleted bands in the borders and the reduction of the width of $\alpha_2$ lamellae, as well as some tungsten segregation in these precipitates. However, because of its low diffusivity, the tungsten content in the $\gamma$ phase is maintained at a sufficient level close to 2 at. %, which sufficiently preserves the mechanical properties at high temperatures.

This particularly good behaviour of the IRIS alloy after exposure as compared to many other TiAl alloys, actually results from the initial homogeneous repartition of tungsten atoms, owing to the use of the SPS route and to the low diffusivity of this element, thus limiting segregations which would be detrimental for the alloy strength. Another advantage of the very weak diffusivity of tungsten is to allow a strong hardening effect for a content of only 2 at. % in the alloy chemical composition, namely at a level which prevents the formation of $\beta_0$ grains. The latter would result in an alloy softening at high temperatures. However, the minimum amount of tungsten required to sufficiently harden the alloy is not known. Alternatively, the current investigations are focused on the stabilisation of the lamellar microstructure through the incorporation of elements like silicon or carbon and/or a slight decrease in the aluminium content, which can also represent interesting solutions for an even better resistance of the IRIS-SPS alloy during ageing.

*Acknowledgements*

This study has been (partially) supported through the grant NanoX n° ANR-17-EURE-0009 in the framework of the « Programme des Investissements d'Avenir" through the project "ALTIAUTO".